\documentclass[journal]{IEEEtran}

\usepackage{amsmath,amssymb,amsfonts,amsthm,mathtools}

\usepackage[utf8]{inputenc}
\usepackage[T1]{fontenc}
\usepackage{hyperref}
\usepackage{url}
\usepackage{booktabs}
\usepackage{nicefrac}
\usepackage{microtype}
\usepackage{xcolor}
\usepackage{graphicx}
\usepackage{subcaption}
\usepackage{multirow}
\usepackage{cite}
\usepackage{seqsplit} 
\usepackage{orcidlink}

\usepackage[capitalize,noabbrev]{cleveref}

\newcommand{\R}{\mathbb{R}}

\newcommand{\zone}{z}                 
\newcommand{\nzones}{Z}                
\newcommand{\sched}{g}                 
\newcommand{\pshort}{p_{\text{short}}} 
\newcommand{\vcurt}{V_{\text{curt}}}   
\newcommand{\voll}{\text{VOLL}}        
\newcommand{\pcurt}{p_{\text{curt}}}   
\newcommand{\Gmax}[1]{G_{\max,#1}}     
\newcommand{\Fmax}[1]{F_{\max,#1}}     

\title{The Operational Value of Spatial Dependence in Renewable Forecast
Scenarios for Single-Period Economic Dispatch: A Controlled Ablation Study}

\author{Jayakumar~Manoharan~\orcidlink{0009-0009-7765-9165},~\IEEEmembership{Senior~Member,~IEEE}%
\thanks{J. Manoharan is with the Electric Power Research Institute (EPRI), Charlotte, NC, USA (e-mail: jmanoharan@epri.com; ORCID: 0009-0009-7765-9165).}}

\begin{document}

\maketitle

\begin{abstract}
Renewable forecasts are evaluated by statistical skill (e.g., CRPS), but
grid operators pay for realized dispatch cost. We diagnose what drives
dispatch value in a single-period newsvendor-style economic dispatch
using real public data from two European transmission systems (CWE,
DE-4TSO). Spatial coherence across forecast sites falls below the
pre-specified 1\% practical-significance threshold: a controlled
ablation holding per-zone marginal forecasts bit-identical and varying
only cross-zone dependence (10 configurations, 3 seeds, paired-bootstrap
confidence intervals) shows a coherence gain of at most 0.64\% of
dispatch cost, indistinguishable from zero in 3 of 10 configurations,
reached only under an unrealistic 8x forecast-error stress test.
Decision-focused training, an established paradigm in this venue,
delivers a robust 2.82-5.19\% gain. A parametric Gaussian-copula
approximation matches the empirical copula at realistic error
magnitudes but performs \emph{worse} than no dependence under extreme
stress. A single-seed sweep: a 12\% energy-score gain changes cost by
less than 0.1\%. Results characterize this single-period dispatch
class; a lightweight four-period extension supports the same
conclusion. For this dispatch class, spatially-correlated scenario generation
provides limited operational value on its own; grid operators and
forecast vendors should instead evaluate dependence models by
downstream decision value and prioritize decision-focused training.

\end{abstract}


\section{Introduction}
\label{sec:introduction}

Renewable forecasts are trained and reported by statistical skill (CRPS),
but grid operators pay for dispatch cost; the two are not the same
target. A large body of work on probabilistic renewable forecasting
optimizes accuracy metrics on the assumption that more accurate forecasts
translate into better operational decisions. A separate, more recent line
of work trains forecasting and dispatch jointly, and often further assumes
that preserving the spatial dependence structure across forecast sites
(so that a wind dip at one plant is correctly modeled as correlated with a
dip at a nearby plant) is a prerequisite for extracting the full decision
value of a forecast. Neither assumption has, to our knowledge, been tested
by an ablation that cleanly isolates the contribution of spatial coherence
from decision-focused training itself, or from the per-zone
marginal-accuracy changes that typically accompany a switch from
independent to joint scenario modeling.

This paper diagnoses what actually drives dispatch value using real
public grid data. Our central result is a strictly controlled ablation
(Experiment B) that holds per-zone marginal forecasts bit-identical by
construction (via within-column permutation of empirical forecast
residuals) and varies only the cross-zone dependence structure (the
copula), isolating the coherence contribution from the confound present
in the most directly comparable concurrent work, whose own
separate-versus-joint ablation changes marginal accuracy and dependence
structure at the same time. \Cref{fig:expb_forest} previews the result:
across 10 configurations, 3 seeds each, the pooled coherence gain never
exceeds 0.64\% of dispatch cost and never reduces curtailment, even under
an unrealistic 8$\times$ forecast-error inflation stress test, while
decision-focused training, an established paradigm in this
venue~\cite{zhang2026dflreview}, delivers a real gain of 2.82-5.19\% in
our experiments.

The main experiments characterize a specific dispatch class: a
single-period newsvendor-style economic dispatch with no unit commitment,
reserve procurement, ramp-rate limits, or temporal coupling across hours.
This is a deliberate scope choice (\cref{sec:method} explains why), and
every claim below should be read with that qualifier attached unless
stated otherwise; \Cref{sec:discussion:richer} reports a lightweight
four-period rolling-horizon sensitivity extension of the central ablation.

\textbf{Contributions.}
\begin{enumerate}
\item A strictly controlled, reusable ablation design isolating spatial
coherence from marginal-accuracy and decision-focus effects: the
coherence contribution is below the pre-specified 1\% practical-
significance threshold (at most 0.64\% of dispatch cost) in this
dispatch class.
\item A decision-focused training gain of 2.82-5.19\% (grid- and
penetration-dependent), robust to high shortage prices and renewable
penetration, individually and jointly.
\item A reconciliation with a near-simultaneous competing result whose
combined decision-focus-and-correlation gain overlaps ours, explaining
why the two numbers are not directly comparable.
\item A supporting, exploratory observation (single seed, descriptive
only): a 12\% energy-score improvement changes mean operating cost by
less than 0.1\%, non-monotonically.
\item A finding that the specific dependence \emph{model}, not just
whether dependence is modeled, can affect dispatch value once the
dependence signal is strong enough: a Gaussian copula matches the
empirical copula at realistic forecast-error magnitudes, but performs
worse than assuming no dependence under an extreme stress test
(\Cref{sec:results:gausscop}).
\end{enumerate}

A substantial and active line of technical work invests engineering
effort in dependency-aware scenario generation and
selection~\cite{tang2017economicdispatch,zhou2026decisionfocused} as a
prerequisite for stochastic dispatch. Our results indicate that, for the
single-period dispatch class studied here, that investment does not pay
off on its own, and, given that a 12\% energy-score improvement here
changes cost by less than 0.1\% (\Cref{sec:results:skillvalue}), that the same budget
is better spent on decision-focused, decision-relevant forecast
improvements rather than on either dependence modeling or marginal
statistical accuracy in isolation.

\Cref{sec:related-work} positions this work against the most directly
related literature; \Cref{sec:method} describes the data, dispatch
formulation, and coherence ablation; \Cref{sec:results} presents
results; \Cref{sec:discussion} reconciles the decision-focused result
against concurrent work; \Cref{sec:limitations} states scope
limitations; \Cref{sec:conclusion} concludes.

\section{Related Work}
\label{sec:related-work}

\paragraph{Probabilistic renewable forecasting.} A large literature
produces probabilistic wind and solar
forecasts~\cite{zhang2014probabilistic,doubleday2021probabilistic},
including copula- and quantile-based methods that already model per-site
predictive spread~\cite{bessa2012quantilecopula}; this paper's forecasting
architecture (\Cref{sec:method:training}) sits within this line of work,
but asks a question this literature does not directly address: once
marginal accuracy is fixed, how much of the resulting decision value comes
from spatial dependence versus from training the forecast against
decision cost directly.

\paragraph{Decision-focused learning for stochastic dispatch.}
Decision-focused (``smart predict-then-optimize'') training, in which a
predictive model is trained end-to-end against downstream decision cost
rather than a standalone accuracy metric,
is by now an established paradigm with foundational task-based
formulations~\cite{elmachtoub2022smart,donti2017taskbased,wilder2019melding},
recent surveys~\cite{kotary2021survey,mandi2024decisionfocused}, and a
review specific to power-system decision-making under
uncertainty~\cite{zhang2026dflreview}. We apply it to grid dispatch; our
contribution is not to establish its value but, via a controlled
copula-only ablation, to isolate the operational value of spatial
dependence when decision-focused training is held fixed. Concurrent with our work, Zhou et al.~\cite{zhou2026decisionfocused} propose
a decision-focused generative-forecasting framework for a two-stage
day-ahead/real-time dispatch with reserve procurement on the IEEE 14-bus
system, reporting 0.80-2.02\% operating-cost reduction over
accuracy-oriented baselines. Their ablation separates
``independently-trained'' from ``jointly-trained'' (correlated) generative
forecasts under fixed decision-focused training, attributing an additional
0.34-1.66 percentage points to joint or correlated forecasting. This
comparison is not a clean copula-only ablation: training a joint generative
model can also change each bus's own marginal forecast accuracy, so their
correlation-associated effect is confounded with marginal-accuracy
improvement. Our \Cref{sec:method:expb} instead holds marginals
bit-identical and varies only the dependence structure, and finds that in
our single-period newsvendor dispatch this clean copula-only effect is
bounded at 0.64\% pooled even under an unrealistic 8$\times$ forecast-error
stress test, and never reduces curtailment. The two results are complementary
rather than contradictory: they suggest the value of spatial correlation is
formulation-dependent, and that disentangling it from confounded
marginal-accuracy gains, as our ablation does, requires a cleaner design
than a change of generative-model architecture provides. Our own
decision-focused result (2.82-5.19\% cost reduction from decision-focused
training alone, \Cref{sec:results}) is likewise not directly comparable to
their combined number; we return to this magnitude question in
\Cref{sec:discussion}.

\paragraph{Value-oriented renewable forecasting.}
Zhang et al.~\cite{zhang2023valueoriented} establish that standard
forecast-accuracy training overlooks the operational value of forecasts and
propose an iterative, value-oriented training alternative that achieves
lower dispatch cost than accuracy-trained forecasts on a virtual-power-plant
dispatch problem. This is a \emph{solution} paper: it proposes a fix and
demonstrates it works. Our skill-value gap result (\Cref{sec:results}) is
instead a \emph{diagnostic} finding under the current, accuracy-oriented
training regime: holding training fixed, we show a 12\% energy-score
improvement yields less than 0.1\% cost improvement, non-monotonically. Zhang et
al.\ already establish that accuracy is not the same as value as a general
principle, which narrows, without eliminating, the novelty of our specific
diagnostic curve. Wen and Pinson~\cite{wenpinson2025valueoriented} pursue a
related but distinct value-oriented question, reconciling forecast
hierarchies across multiple market participants with different objectives
via a cooperative-game formulation; their setting (multi-agent forecast
reconciliation for wind energy trading) does not overlap with our
single-operator dispatch setting, but their framing that accuracy-optimal
reconciliation need not be value-optimal for every participant is
consistent with the broader pattern we diagnose. Earlier work already
showed that using forecast uncertainty, not just a point forecast, changes
the economics of wind participation in
markets~\cite{pinson2007trading,morales2009economic}; our skill-value
result extends this line to ask specifically about the value of
\emph{spatial} forecast structure, and a fully differentiable
multistage decision-focused forecasting
architecture~\cite{persak2024decisionfocused} is a close methodological
neighbor to our decision-focused training regime.

\paragraph{Forecast skill versus decision skill, more generally.}
Raeth and Ludwig~\cite{raethludwig2025forecastskill} establish empirically,
across general weather-dependent decision tasks that are not specific to
grid dispatch, that forecast-level skill does not reliably translate into
downstream decision performance. This is the general principle behind our
skill-value gap result; our contribution narrows it to the specific,
grid-dispatch instantiation and its non-monotonic energy-score-versus-cost curve,
rather than establishing the general accuracy-does-not-equal-value claim
itself.

\paragraph{Copula-based and spatially-correlated scenario generation.}
Tang et al.~\cite{tang2017economicdispatch} model spatial and temporal
correlation among multiple renewable plants using copula theory and a
Gibbs-sampling scenario generator, and show that accounting for these
correlations changes the solution of a real-time economic dispatch problem
relative to treating plants independently. This sits within a broader
literature on preserving cross-site
dependence when generating renewable
scenarios~\cite{papaefthymiou2008spatialdependence,pinson2009scenarios} and
more recent data-driven alternatives such as generative adversarial
networks~\cite{chen2018gan}; separately, the spatial and
temporal error structure this machinery is built to capture has itself
been characterized
directly~\cite{fang2018modelling,bruninx2014statistical}. This literature
motivates the question our \Cref{sec:method:expb} answers directly: once
decision-focused training and marginal forecast accuracy are held fixed,
how much value does the correlation structure itself contribute? We are
not aware of a prior ablation that isolates this quantity as cleanly as
ours, which is why we present it as a reusable methodological contribution
in addition to an empirical finding.

\paragraph{Stochastic and robust dispatch under renewable uncertainty.}
A separate, large literature manages renewable uncertainty directly inside
the dispatch or unit-commitment formulation, via chance-constrained and
scenario-based stochastic
programming~\cite{wang2012chanceconstrained}
and distributionally robust or adaptive robust
reformulations~\cite{bertsimas2013robust}.
These formulations are largely orthogonal to our question: they ask how to
dispatch \emph{given} an uncertainty representation, whereas we ask how
much of that representation's spatial-dependence component is worth
modeling in the first place, holding the dispatch formulation itself
fixed. Our controlled ablation design is compatible with, and could in
principle be applied on top of, any of these richer formulations.

\section{Method}
\label{sec:method}

\subsection{Data and grids}
\label{sec:method:data}

We use real, publicly available data from Open Power System
Data~\cite{wiese2018openpower} throughout; no synthetic or smoke-test
numbers appear in this paper. We report results on two European grids: a
9-zone Central-West-Europe (CWE) grid (Germany, France, the Netherlands,
Belgium, Denmark, Austria, Switzerland, Czechia, and Poland,
interconnector capacity fraction $\alpha_{\text{ntc}}=0.10$, the fraction
of each line's rated thermal capacity assumed available as net transfer
capacity), and a 4-zone
German transmission-system-operator (DE-4TSO) grid, pooling 7535 and 7557
test hours respectively per seed for the decision-focused penetration
result (labeled \textbf{C1} in \Cref{tab:zhou_comparison}, the Discussion,
and the appendix's claim-file mapping) and the skill-value-gap result
(labeled \textbf{C2} in the same locations). The CWE grid was
introduced specifically because its interconnectors bind: at baseline
(1.0$\times$) penetration we confirm directly that transmission lines are
at their capacity limit in 67-69\% of test hours, with the DE-FR,
DE-NL, DE-CH, and DE-PL lines binding in nearly 100\% of hours (the
fraction drops modestly, to 65-68\%, at the higher penetration levels
swept in \Cref{sec:results:c1}), ruling out the possibility that any
result below is an artifact of a grid whose limits never bind.

\subsection{Dispatch formulation}
\label{sec:method:dispatch}

The main results use a single-period newsvendor-style economic dispatch:
a here-and-now schedule $\sched_0$ is chosen once against a scenario
ensemble of size $S=20$, and the realized cost is evaluated against the
true outcome via curtailment and shortage penalties, with \emph{no} unit
commitment, reserve procurement, ramp-rate limits, or temporal coupling
across hours ($H=1$). This is a deliberate scope decision: a full
two-stage stochastic unit commitment with binary commitment variables
would delete the envelope-theorem gradient our decision-focused training
relies on to back-propagate through the dispatch layer, and cost two to
three orders of magnitude more per solve (\Cref{sec:discussion:richer}
reports a lightweight multi-period extension that does not train this
differentiable objective and so is not constrained by this tradeoff).
Every claim in the main results characterizes this dispatch class
specifically, except for that extension; \Cref{sec:limitations} returns to
what this scope excludes.

Network coupling between zones enters only through explicit bilateral
line-flow variables bounded by net-transfer-capacity (NTC) limits: a
deliberately simplified zonal transportation-style exchange model, not
the Flow-Based Market Coupling used in the Core capacity calculation
region, nor a nodal admittance-based DC power flow with Kirchhoff's
voltage law (no per-node price decomposition).
\Cref{sec:method:data}'s interconnector binding statistics report how
often these NTC limits actually constrain the zonal dispatch.

\textbf{Base dispatch QP.} Every dispatch solve in this paper, whether
evaluating a committed schedule (\Cref{sec:method:training}) or a scenario
within a stochastic recourse problem (\Cref{sec:method:expb}), reduces to
the same single-scenario quadratic program. Given a renewable output
vector $\hat{r}\in\R^{\nzones}$ (a committed value, a scenario draw, or a
realized outcome) and load $\ell\in\R^{\nzones}$, define
\begin{subequations}
\label{eq:base-qp}
\begin{align}
\mathrm{Dispatch}(\hat{r},\ell) \;=\!\! \min_{g,\mathrm{curt},\mathrm{shed},f} &
\sum_{\zone=1}^{\nzones}\!\left(a_\zone g_\zone^2 \!+\! b_\zone g_\zone\right) \nonumber\\
& +\, \vcurt\!\sum_{\zone=1}^{\nzones}\!\mathrm{curt}_\zone
+ \voll\!\sum_{\zone=1}^{\nzones}\!\mathrm{shed}_\zone \label{eq:base-qp:obj}\\
\text{s.t.}\ \
& g_\zone - \mathrm{curt}_\zone + \textstyle\sum_{l=1}^{L} A_{\zone l} f_l + \mathrm{shed}_\zone \nonumber\\
&\hspace{3.5em} = \ell_\zone - \hat{r}_\zone, \quad \forall \zone \label{eq:base-qp:bal}\\
& 0 \le g_\zone \le \Gmax{\zone}, \quad \forall \zone \label{eq:base-qp:gmax}\\
& 0 \le \mathrm{curt}_\zone \le \hat{r}_\zone, \quad \forall \zone \label{eq:base-qp:curt}\\
& \mathrm{shed}_\zone \ge 0, \quad \forall \zone \label{eq:base-qp:shed}\\
& -\Fmax{l} \le f_l \le \Fmax{l}, \ \forall l, \label{eq:base-qp:flow}
\end{align}
\end{subequations}
where $g_\zone\ge0$ is thermal generation (quadratic merit-order cost),
$\mathrm{curt}_\zone$ is curtailed renewable output (penalized at
$\vcurt=200$), $\mathrm{shed}_\zone$ is unserved load (penalized at
$\voll=3000$), $f_l$ is the bilateral flow on line $l$ bounded by its NTC
limit $\Fmax{l}$, and $A\in\{-1,0,1\}^{\nzones\times L}$ is the network
incidence matrix ($A_{\zone l}=+1$ if line $l$ delivers into zone $\zone$,
$-1$ if it leaves, $0$ otherwise), so $\sum_l A_{\zone l}f_l$ is zone
$\zone$'s net import. Cost figures are in the simulator's internal cost
units (a linear-quadratic merit-order model calibrated for relative, not
absolute, realism); we report percentage cost reductions throughout so
results do not depend on this calibration.

\subsection{Decision-focused versus accuracy-oriented training}
\label{sec:method:training}

We train the same forecasting architecture (a neural scenario generator
producing a per-zone predictive mean $\mu\in\R^{\nzones}$ and standard
deviation $\sigma\in\R^{\nzones}$ from input features; see the Data and
Code Availability paragraph, \Cref{sec:conclusion}, for implementation
details) under two objectives. Given $\mu,\sigma$, the
committed renewable schedule is the newsvendor-style safety margin
\begin{equation}
c_\zone = \max\!\big(0,\ \mu_\zone - \kappa\sigma_\zone\big), \qquad \zone=1,\dots,\nzones,
\label{eq:commit}
\end{equation}
with $\kappa=1$ throughout, and the committed dispatch cost is
$C_{\mathrm{commit}} = \mathrm{Dispatch}(c,\ell)$ (\Cref{eq:base-qp}),
evaluated against the true load $\ell$: we assume load is known exactly
and only renewable output is uncertain, consistent with load forecasts
being far more accurate than renewable forecasts at the horizons studied
here. Writing $r\in\R^{\nzones}$ for the true realized renewable output and
$\delta = r - c$ for the realized imbalance, the total realized cost is
\begin{equation}
\mathrm{Cost} = C_{\mathrm{commit}} + \pcurt\!\sum_{\zone=1}^{\nzones}\!\max(0,\delta_\zone) + \pshort\!\sum_{\zone=1}^{\nzones}\!\max(0,-\delta_\zone),
\label{eq:realized-cost}
\end{equation}
with $\pcurt=5$ and $\pshort=80$ (swept to 160 in \Cref{sec:results:c1}).
This realized-imbalance penalty is a separate, second cost layer from
$\vcurt,\voll$ already inside $\mathrm{Dispatch}$ (\Cref{eq:base-qp}),
which apply only to the smaller, distinct effect of the committed
schedule's \emph{own} dispatch curtailing or shedding relative to $c$.
Because $\pshort \gg \pcurt$, the cost-optimal commitment is a
conservative \emph{quantile} of the predictive distribution rather than
its mean, the newsvendor structure referenced throughout. Both
$C_{\mathrm{commit}}$ (via the envelope theorem: the QP's optimal value's
gradient with respect to \Cref{eq:base-qp:bal}'s right-hand side equals
that constraint's dual multiplier, computed once per forward solve and
reused on the backward pass) and \Cref{eq:commit,eq:realized-cost} are
differentiable in $\mu,\sigma$, so \Cref{eq:realized-cost} back-propagates
directly into the scenario generator's parameters, in the spirit of
differentiable optimization layers for end-to-end
learning~\cite{amoskolter2017optnet}.

The \emph{accuracy-oriented} (``crps'') regime minimizes the energy score
(ES), the multivariate generalization of the univariate continuous ranked
probability score (CRPS), instead, never seeing \Cref{eq:realized-cost}
during training; the \emph{decision-focused} (``cost'') regime trains
directly on it. Comparing the realized cost of
schedules from these two regimes, holding the grid, penetration, and
ensemble fixed, is the decision-focused value gain (\Cref{sec:results:c1}).
A $\lambda$-weighted regularizer, loss $=\mathrm{Cost} + \lambda\cdot
\mathrm{ES}(\text{scenarios}, r)$, interpolates between the two regimes,
tracing the energy-score-versus-cost curve of \Cref{sec:results:skillvalue}.

\subsection{Experiment B: an isolated coherence ablation}
\label{sec:method:expb}

Our central methodological contribution isolates the value of spatial
coherence from both decision-focus and marginal-accuracy effects. We use
persistence-based marginal forecasts ($\mu_t = r_{t-1}$) so that no learned
forecasting model is involved, and construct the predictive spread from
the empirical persistence-residual pool estimated on the time-ordered
training split. For each test instance we draw $S$ residual rows to build
two scenario ensembles: a \textsc{Real} ensemble, where residual rows are
drawn intact and preserve real cross-zone dependence, and an
\textsc{Indep} ensemble, where each zone's column is independently permuted,
destroying cross-zone dependence while leaving each zone's own marginal
distribution untouched. Because \textsc{Indep} permutes \emph{within}
columns of the same residual matrix, the two ensembles have bit-identical
per-zone marginals by construction, asserted numerically for every test
instance; only the copula differs.

Both ensembles feed the same two-stage stochastic dispatch with
network-coupled recourse. Given a scenario ensemble
$\{r^{(s)}\}_{s=1}^{S}$ (either \textsc{Real}, \textsc{Indep}, or the
\textsc{Gauss} arm of \Cref{sec:results:gausscop}) and load $\ell$, the
here-and-now schedule $\sched_0\ge0$ solves
\begin{subequations}
\label{eq:two-stage}
\begin{align}
\min_{\sched_0} \ & \sum_{\zone=1}^{\nzones}\!\left(a_\zone \sched_{0,\zone}^2 \!+\! b_\zone \sched_{0,\zone}\right) \nonumber\\
& +\, \frac{1}{S}\sum_{s=1}^{S} \Big[\pshort\!\sum_{\zone}\! u_\zone^{(s)} \!+\! \vcurt\!\sum_{\zone}\!\mathrm{curt}_\zone^{(s)} \!+\! \voll\!\sum_{\zone}\!\mathrm{shed}_\zone^{(s)}\Big] \label{eq:two-stage:obj}\\
\text{s.t.}\ \
& \sched_{0,\zone} + u_\zone^{(s)} \!-\! d_\zone^{(s)} \!-\! \mathrm{curt}_\zone^{(s)} \!+\! \textstyle\sum_l A_{\zone l} f_l^{(s)} \!+\! \mathrm{shed}_\zone^{(s)} \nonumber\\
&\hspace{3.7em} = \ell_\zone - r_\zone^{(s)}, \quad \forall \zone,s \label{eq:two-stage:bal}\\
& 0 \le \sched_{0,\zone} + u_\zone^{(s)} - d_\zone^{(s)} \le \Gmax{\zone}, \quad \forall \zone,s \label{eq:two-stage:gmax}\\
& 0 \le \mathrm{curt}_\zone^{(s)} \le r_\zone^{(s)}, \ u_\zone^{(s)},d_\zone^{(s)},\mathrm{shed}_\zone^{(s)} \ge 0, \ \forall \zone,s \label{eq:two-stage:nonneg}\\
& -\Fmax{l} \le f_l^{(s)} \le \Fmax{l}, \ \forall l,s, \label{eq:two-stage:flow}
\end{align}
\end{subequations}
where $u^{(s)},d^{(s)}\ge0$ are per-scenario upward/downward recourse
around $\sched_0$: $u^{(s)}$ (covering a shortfall above $\sched_0$) is
penalized at $\pshort$, a proxy for fast balancing generation, while
$d^{(s)}$ (reducing output below $\sched_0$) is free, since curtailing
owned thermal capacity is costless here; only $\mathrm{curt}^{(s)}$ and
$\mathrm{shed}^{(s)}$ carry the same $\vcurt,\voll$ costs as
\Cref{eq:base-qp}. The realized cost of a committed $\sched_0$ against the
true outcome $r$, $\mathrm{RealizedCost}(\sched_0, r,\ell)$, is the
single-scenario ($S{=}1$) specialization of \Cref{eq:two-stage} with
$\sched_0$ fixed. Writing $C_{\textsc{Real}}$ and $C_{\textsc{Indep}}$ for
the mean realized cost of the schedules chosen by solving
\Cref{eq:two-stage} under each ensemble, the coherence gain is
\begin{equation}
\text{gain} = 100 \times \frac{C_{\textsc{Indep}} - C_{\textsc{Real}}}{C_{\textsc{Indep}}},
\label{eq:coherence-gain}
\end{equation}
so that a positive gain means that knowing the empirical historical
cross-zone dependence structure (as in \textsc{Real}) reduces realized cost relative to
discarding it (as in \textsc{Indep}), expressed as a percentage of the
\textsc{Indep} cost.

Before running this experiment on real historical residuals, we
pre-specified a decision rule for interpreting its outcome, based on a
prior scoping analysis with synthetic, deliberately extreme cross-zone
copulas that found a coherence effect around 4\%. Real inter-zone
forecast-error correlations are typically much weaker (Pearson
$\sim$0.1-0.3), so we expected the real-data effect to plausibly fall
below 1\%, and fixed the interpretation in advance as a gain of at least
1\% of dispatch cost with a curtailment reduction of at least 0.5
percentage points, versus a smaller effect on either leg.

We evaluate 10 configurations. Seven vary renewable penetration:
1.0$\times$, 1.5$\times$, 2.0$\times$ (clipped and uncapped), 3.0$\times$
(clipped and uncapped), and 4.0$\times$ (uncapped). The remaining three
hold penetration at 3.0$\times$, uncapped, and inflate the forecast-error
magnitude applied to the residual pool by 2$\times$, 4$\times$, and
8$\times$ its historically-estimated value, as a stress test of how large
forecast errors would need to be before coherence starts to matter. The
seven penetration-sweep configurations pool 300 test instances per seed;
the three stress configurations pool 200. Every configuration runs with 3
random seeds; \Cref{sec:results} reports pooled, paired-bootstrap CI
estimates alongside per-seed point estimates, where ``pooled'' means the
per-instance costs are combined across all seeds before computing the
gain of \Cref{eq:coherence-gain} and its CI, rather than averaging
per-seed gains (numerically equivalent for the point estimate, but the
pooled bootstrap uses the full per-instance sample for the interval).

\section{Results}
\label{sec:results}

\begin{figure*}[t]
    \centering
    \includegraphics[width=0.78\textwidth]{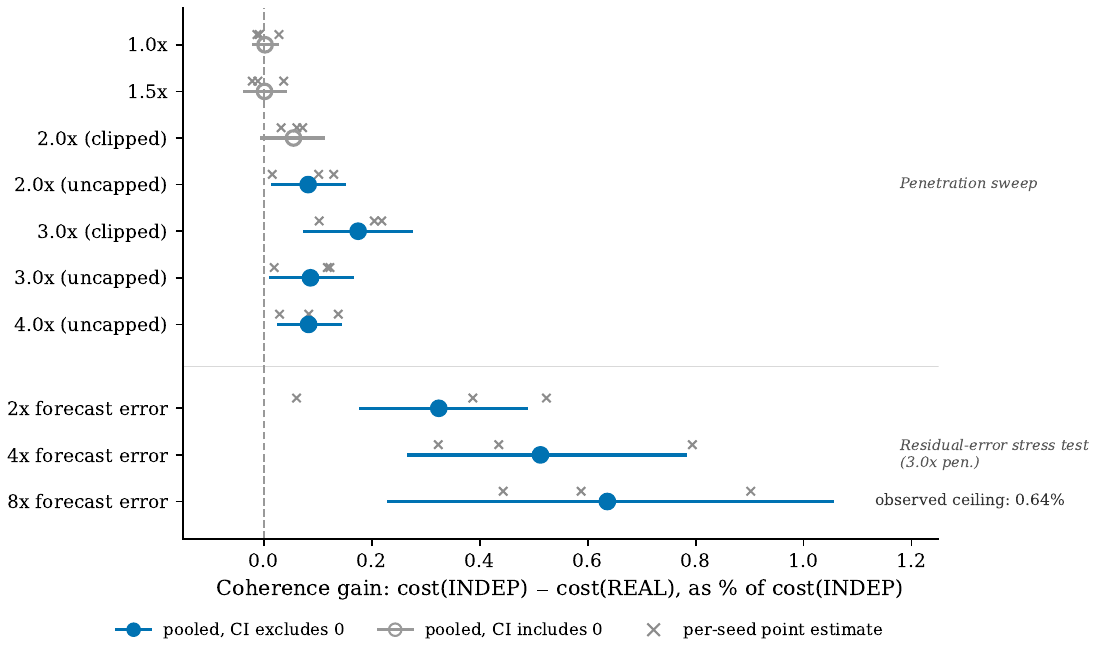}
    \caption{Coherence-ablation forest plot, isolating the effect of the
    cross-zone dependence structure (the ``copula'') on dispatch cost while
    holding marginal forecast accuracy bit-identical by construction. Top
    group varies renewable penetration (1.0$\times$-4.0$\times$, clipped
    and uncapped surplus handling); bottom group fixes penetration at
    3.0$\times$ and inflates forecast-error magnitude (2$\times$-8$\times$)
    as a stress test. Filled blue circles with solid whiskers: 95\%
    paired-bootstrap CI (3-seed pooled) excludes zero; open gray circles:
    CI includes zero; small crosses: the three per-seed point estimates.
    Even at the observed pooled ceiling (8$\times$ inflation, 0.64\%), the
    effect stays below the $\ge$1\% practical-significance threshold, and
    curtailment is unchanged to within $5\times10^{-4}$ percentage points in
    every configuration (not shown). Source-file mapping: Appendix~A.}
    \label{fig:expb_forest}
\end{figure*}
\begin{figure*}[t]
    \centering
    \includegraphics[width=0.72\textwidth]{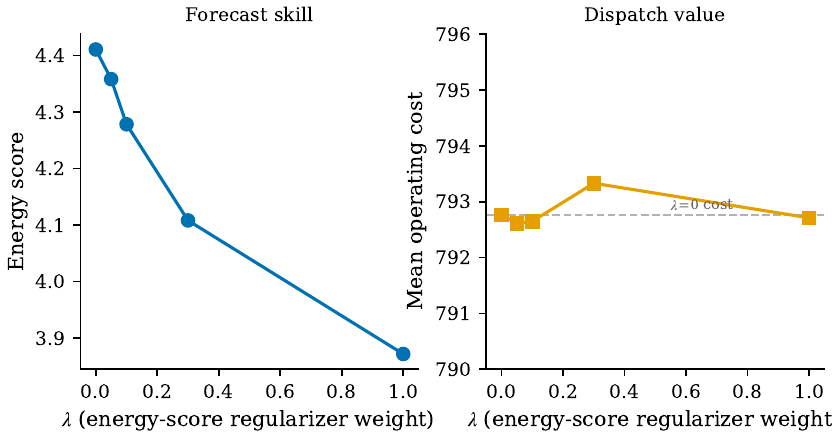}
    \caption{A 12\% energy-score improvement (left panel, 4.41$\to$3.87 as $\lambda$
    (the weight on an energy-score regularizer added to decision-focused training)
    increases from 0 to 1.0, DE-4TSO, seed 0) changes mean operating cost by
    less than 0.1\% with no systematic trend (right panel; dashed line marks
    the $\lambda=0$ reference cost). Shown as two panels sharing a $\lambda$
    axis rather than a dual-axis overlay, since the two series' relative
    scales (12\% vs.\ 0.09\%) would otherwise create a misleading visual
    coupling. Source-file mapping: Appendix~A.}
    \label{fig:c2_skillvalue}
\end{figure*}
\begin{figure*}[t]
    \centering
    \includegraphics[width=0.74\textwidth]{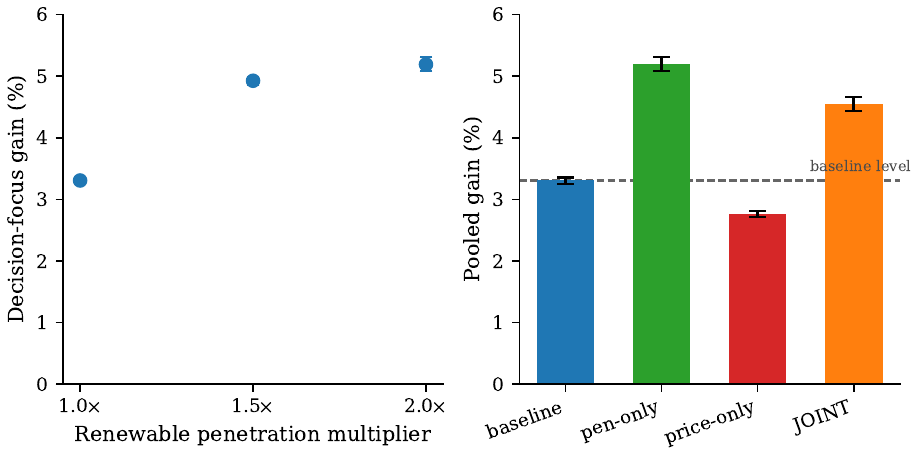}
    \caption{\textbf{Left:} decision-focused value gain (\% reduction in mean
    operating cost relative to accuracy-oriented training, same forecast
    inputs) vs. renewable penetration (CWE grid, uncapped surplus): 3.31\%
    (1.0$\times$, 3-seed pooled) $\to$ 4.93\% (1.5$\times$, 3-seed pooled)
    $\to$ 5.19\% (2.0$\times$, 3-seed pooled). Points
    are not connected, since only three discrete penetration levels were
    tested. \textbf{Right:} joint robustness to two independent stresses:
    the same decision-focus gain under baseline, high penetration alone,
    high shortage price alone, and both simultaneously (JOINT), with the
    baseline level marked (dashed line). The gain exceeds baseline under high
    penetration and under the joint adverse condition, and is not the
    smallest of the four corners even when both stresses apply
    simultaneously. Error bars are 95\% paired-bootstrap CIs. Source-file
    mapping: Appendix~A.}
    \label{fig:c1_robustness}
\end{figure*}


\subsection{Isolated value of empirical spatial coherence}
\label{sec:results:expb}

\Cref{fig:expb_forest} shows the coherence gain, as defined in
\Cref{sec:method:expb}, across all 10 configurations, pooling 3 random
seeds per configuration with 95\% paired-bootstrap confidence intervals,
and overlaying the three individual per-seed point estimates alongside
each pooled estimate. Seven of the ten configurations have confidence
intervals that exclude zero, meaning the coherence effect is
statistically detectable; however, every 3-seed pooled point estimate
remains at or below 0.64\% of dispatch cost, and this ceiling is reached
only under an unrealistic 8$\times$ forecast-error inflation stress test
applied on top of 3$\times$ renewable penetration. Individual per-seed
point estimates are noisier than the pooled ceiling: the single largest,
one seed of that same 8$\times$ configuration, reaches 0.90\%, still below
the $\ge$1\% practical-significance threshold. Under realistic
forecast-error magnitudes (the 1$\times$ residual-scale rows), the effect
never exceeds 0.5\% pooled. Across all configurations and seeds, realized
curtailment differs by less than $5\times10^{-4}$ percentage points
between the \textsc{Real} and \textsc{Indep} copula arms, so the coherence
effect never translates into a measurable curtailment reduction.

Since the mechanism above ties the effect to congestion, we re-ran the
baseline, two uncapped-penetration points, and the ceiling configuration
at $\alpha_{\text{ntc}}\in\{0.05,0.15,0.20\}$ against the $0.10$ default
(Appendix~A): near-zero configurations stay near zero throughout
($\le0.14\%$); the ceiling gain does not grow as congestion tightens,
instead varying non-monotonically between $-0.06\%$ and $0.64\%$, with
95\% CIs including zero at two of four values. Tighter congestion does
not unlock a larger coherence value.

A plausible (not proven) mechanistic explanation: the here-and-now
schedule solves a stochastic QP whose expected recourse cost is estimated
from $S=20$ scenarios per instance. With bit-identical marginals, the two
copula arms differ only in how the joint tail of cross-zone outcomes is
represented within those 20 draws; when interconnectors bind, this joint
tail matters to the network-coupled recourse cost, but with only 20
scenarios it is coarsely resolved, so its contribution is small relative
to the identical marginal contribution. This predicts the coherence
effect could grow with a finer ensemble, taken up directly in
\Cref{sec:results:scenariocount}.

The seed-to-seed spread is itself informative: point estimates vary by up
to $9\times$ across seeds within a single configuration (e.g., at
3$\times$ penetration with 2$\times$ residual-error inflation, the three
seeds give 0.06\%, 0.52\%, and 0.39\%), indicating the per-instance effect
is not a stable, exploitable signal even where the pooled CI excludes
zero.

\subsection{Sensitivity to scenario count}
\label{sec:results:scenariocount}

The mechanistic explanation above predicts that a coarsely-resolved
scenario ensemble ($S=20$) could itself be why the coherence effect stays
small: with more scenarios, the joint tail is resolved more finely, and
the copula arms might separate further. We tested this by re-running four
of the ten configurations (baseline penetration, $3\times$ penetration,
and the $3\times$-penetration $4\times$/$8\times$ forecast-error stress
tests, including the 0.64\%-ceiling configuration) at $S=20,50,100,200$,
single seed: baseline moves $0.03\%\to0.03\%\to0.03\%\to0.02\%$;
$3\times$ penetration moves $0.08\%\to0.12\%\to0.11\%\to0.12\%$; the
$4\times$-stress configuration moves $0.32\%\to0.06\%\to0.16\%\to0.26\%$;
and the $8\times$-stress ceiling configuration moves
$0.59\%\to0.16\%\to0.34\%\to0.21\%$ (all single-seed point estimates,
$S=20,50,100,200$ respectively). We then repeated the baseline and
ceiling configurations, the two most consequential for this question, at
the same four $S$ values pooled across 3 seeds each, matching the
statistical power of the paper's other headline claims rather than relying
on the single-seed check alone: baseline moves $0.00\%\to0.03\%\to0.02\%
\to0.02\%$ (95\% CIs all include or barely exclude zero at every $S$), and
the ceiling configuration moves $0.64\%\to0.47\%\to0.34\%\to0.21\%$ (95\%
CI $[0.23,1.05]$ at $S=20$ narrowing and shifting down to $[-0.01,0.43]$ at
$S=200$). If anything, the pooled ceiling estimate \emph{decreases}
monotonically as the scenario ensemble grows finer, the opposite of what an
under-resolution artifact would predict; by $S=200$ its confidence
interval includes zero. Every value at every tested $S$, pooled or
single-seed, remains well below the $\ge$1\% threshold, with no trend
toward the ceiling growing as $S$ increases: the single-seed fluctuation
looks like the same per-instance sampling noise documented in
\Cref{sec:results:expb}, and the better-powered pooled sweep shows the
ceiling shrinking, not growing. This does not rule out a qualitatively
different result at much larger $S$, but it removes $S=20$ being simply
"too small" as the explanation, and if anything suggests $S=20$ may
modestly overstate the effect relative to a finer ensemble.

\subsection{Sensitivity to dependence-model specification}
\label{sec:results:gausscop}

\textsc{Real} preserves whatever empirical historical cross-zone dependence
actually occurred, including any non-Gaussian tail dependence; it is not
necessarily the dependence structure a sophisticated, weather-conditioned
forecasting system would produce. We therefore added a third ensemble,
\textsc{Gauss}, that fits a parametric Gaussian copula to the same
residual pool's rank correlation and transforms it through each zone's own
empirical marginal, so it shares \textsc{Real}'s per-zone marginals and
approximately its linear correlation structure, but not its higher-order
(tail) dependence. We evaluate \textsc{Gauss} at four configurations,
pooled across 3 seeds each ($n=900$ for the three realistic-error-scale
penetration levels, $n=600$ for the $8\times$-stress ceiling): baseline
($1.0\times$), $2.0\times$, and $3.0\times$ penetration at the real,
historically-estimated forecast-error scale, and the $3.0\times$
penetration, $8\times$-forecast-error-scale ceiling configuration from
\Cref{sec:results:expb}.

At realistic forecast-error magnitudes, \textsc{Gauss} is
indistinguishable from \textsc{Real}: at baseline, \textsc{Real} beats
\textsc{Indep} by $0.05\%$ (95\% CI $[0.03,0.07]$) and \textsc{Gauss} beats
\textsc{Indep} by a statistically indistinguishable $0.05\%$ (95\% CI
$[-0.01,0.10]$), with \textsc{Real} vs.\ \textsc{Gauss} itself at $0.00\%$
(95\% CI $[-0.05,0.05]$, includes zero); the same pattern holds at
$2.0\times$ ($0.09\%$ vs.\ $0.12\%$, \textsc{Real}-vs-\textsc{Gauss} CI
$[-0.17,0.12]$) and $3.0\times$ ($0.16\%$ vs.\ $0.16\%$,
\textsc{Real}-vs-\textsc{Gauss} CI $[-0.18,0.19]$) penetration. Only under
the $8\times$-forecast-error stress test does the picture change
qualitatively: \textsc{Real} beats \textsc{Indep} by $0.65\%$ (95\% CI
$[0.22,1.09]$, consistent with the pooled value reported in
\Cref{sec:results:expb}), but \textsc{Gauss} \emph{loses} to \textsc{Indep}
by $-0.90\%$ (95\% CI $[-1.78,-0.03]$): a smooth Gaussian copula matched to
the real correlation matrix performs \emph{worse} than assuming no
dependence at all. \textsc{Real} beats \textsc{Gauss} by $1.54\%$ (95\% CI
$[0.68,2.36]$) at this one configuration, a larger and more clearly
significant gap than \textsc{Real} vs.\ \textsc{Indep} itself, and the only
one of the four tested configurations where \textsc{Real} vs.\
\textsc{Gauss} excludes zero. We do not have a fully verified explanation for this, but a plausible one
is that Gaussian copulas have zero tail dependence by construction; at
realistic error magnitudes the joint tail this misses is too small to
matter economically, but under extreme stress, a schedule optimized
against a Gaussian approximation of the real joint tail becomes
miscalibrated for the simultaneous-shortfall scenarios that matter most
once interconnectors bind, while \textsc{Indep} at least does not
pretend to know a (wrong) tail structure.

This is a scoped, not general, finding: at realistic magnitudes the
choice between empirical and Gaussian-parametric dependence has no
detectable consequence, and both modestly outperform no dependence;
only under stress far beyond historical scale does misspecification
become costlier than ignoring dependence. Read together, this paper's
evidence on spatial dependence is threefold: empirical coherence adds
little value once marginals are controlled; finer scenario resolution
does not recover a larger effect; and the wrong parametric dependence
form can make outcomes worse than ignoring dependence once the signal is
strong enough to matter.

\subsection{Statistical forecast skill and dispatch value are decoupled}
\label{sec:results:skillvalue}

\Cref{fig:c2_skillvalue} sweeps a regularizer weight $\lambda$ that
interpolates between accuracy-oriented and decision-focused training
(\Cref{sec:method:training}) on the DE-4TSO grid. As $\lambda$ increases
from 0 to 1.0, the energy score improves by 12\% (4.41 to 3.87), while mean operating
cost changes by less than 0.1\% with no systematic trend as $\lambda$
varies: it is non-monotonic, moving from 792.76 at $\lambda=0$ down to
792.61-792.65 at intermediate $\lambda$, up to 793.33 at $\lambda=0.3$,
and back down to 792.71 at $\lambda=1.0$. A substantial, monotonic
improvement in forecast accuracy therefore produces no corresponding,
systematic improvement in dispatch cost.

\subsection{Decision-focused training buys real, robust value}
\label{sec:results:c1}

On the smaller DE-4TSO grid (today's penetration, no stress conditions),
decision-focused training reduces cost by 2.82\% $\pm$ 0.15 (3-seed
pooled, 95\% CI $[2.65,3.02]$); the value-trained model achieves this
despite a \emph{worse} energy score (+18.5\%) than the accuracy-trained model, which
confirms that decision-focused training and forecast-accuracy training
pursue genuinely different objectives. \Cref{fig:c1_robustness}
(left) reports the same gain on the larger, more congested CWE grid as
renewable penetration scales from 1.0$\times$ to 2.0$\times$: 3.31\%
(1.0$\times$, 3-seed pooled, 95\% CI $[3.26,3.36]$), 4.93\%
(1.5$\times$, 3-seed pooled, 95\% CI $[4.86,4.99]$), and 5.19\%
(2.0$\times$, 3-seed pooled, 95\% CI $[5.08,5.31]$). The 1.0$\times$/shortage-price-80 condition is
also the baseline condition for the price-asymmetry sweep below; the two
sweeps share this single baseline run, which is why the same 3.31\%
figure appears in both.

Doubling the shortage price ($\pshort$, \Cref{sec:method:dispatch}) from
80 to 160 shrinks the gain from 3.31\% to
2.76\% (3-seed pooled, 95\% CI $[2.71,2.81]$): the decision-focused
advantage erodes, but does not disappear, in the high-shortage-cost regime
that matters most operationally. \Cref{fig:c1_robustness} (right) reports
a joint test of both adverse conditions at once: elevated penetration
(2.0$\times$, uncapped) and elevated shortage price (160) simultaneously.
The gain under this joint adverse condition is 4.55\% (3-seed pooled, 95\%
CI $[4.44,4.66]$), which exceeds the baseline (3.31\%) and is not the
smallest of the four corners tested even when both stresses apply at
once. The joint effect is close to a simple additive combination of the
two individual adverse effects (4.65\% predicted vs.\ 4.55\% observed),
indicating a well-behaved interaction rather than a nonlinear collapse.

\section{Discussion}
\label{sec:discussion}

\subsection{Reconciling the magnitude gap with Zhou et al.}
\label{sec:discussion:zhou}

%
\begin{table*}[t]
\centering
\scriptsize
\caption{Setup comparison between this work (C1, Experiment B) and the
near-simultaneous Zhou et al.~\cite{zhou2026decisionfocused}. See the
Discussion section for why these differences make the two papers' headline
numbers not directly comparable.}
\label{tab:zhou_comparison}
\renewcommand{\arraystretch}{0.82}
\begin{tabular}{p{2.6cm}p{6.3cm}p{6.3cm}}
\toprule
\textbf{Dimension} & \textbf{This work (C1 + Experiment B)} & \textbf{Zhou et al.~(2026)} \\
\midrule
Grid &
CWE 9-zone (real European TSO zones, real OPSD data) / DE-4TSO 4-zone &
IEEE 14-bus (11 load buses), real Southern-China load data, proportionally rescaled for feasibility \\
Dispatch formulation &
Single-period newsvendor-style QP: schedule $g_0$ chosen once, then curtailment/shortage penalties only. No commitment, no reserves, no ramp limits, no DC power flow &
Two-stage day-ahead (DA) schedule + reserve capacities, then real-time (RT) recourse redispatch using procured reserves, ramp-limited, DC power flow with line limits \\
Forecast horizon &
$H=1$ (single hour-ahead, persistence-based marginal forecast) &
Day-ahead (multi-hour, $T\approx24$h) with hourly RT recourse \\
Risk framework &
Stochastic average cost over an $S=20$ empirical-residual scenario ensemble &
Distributionally robust optimization (DRO) over an ambiguity set around the generated scenario distribution \\
Price/penalty structure &
Newsvendor critical ratios: shortage price (80 default, tested to 160), curtailment cost 200, value of lost load 3000 &
Reserve deployment costs ($\rho^{\uparrow}/\rho^{\downarrow}$) plus load-shed/curtailment penalties ($\beta_{\text{shed}}/\beta_{\text{cur}}$); no single VOLL-style ratio reported \\
Correlation manipulation &
Experiment B: per-zone marginals held bit-identical by construction (within-column permutation of empirical residuals); only the dependence structure (copula) varies between the two arms &
``Separate'' vs.\ ``Joint'' forecasting: a different generative model architecture is trained in each setting; per-zone marginal accuracy is not held identical between the two arms \\
Reported cost reduction &
2.82-5.19\% (decision-focused training, C1, condition-dependent) &
0.80-2.02\% (combined decision-focus + correlation-aware training) \\
\bottomrule
\end{tabular}
\end{table*}

Our decision-focused-only gain (2.82-5.19\%, depending on grid and
penetration) numerically exceeds Zhou et al.'s combined
decision-focus-and-correlation gain (0.80-2.02\%,
\Cref{sec:related-work}), even though our result is a strict subset of
what their combined framework captures. Table~\ref{tab:zhou_comparison}
lays out the setup differences we believe plausibly inflate our number
relative to theirs, offered as structural explanations rather than a
proof that fully accounts for the gap: our single-period dispatch
concentrates all forecast-error cost sensitivity into one decision, while
their real-time recourse absorbs part of it; their distributionally
robust formulation already buys some robustness our empirical-average
objective does not, exposing the full accuracy-versus-cost-optimal gap;
our newsvendor critical ratios create a more asymmetric cost landscape,
which decision-focused training is designed to exploit; and aggregating
to 4-9 zones rather than 14 buses reduces effective dimensionality,
plausibly widening the training gap relative to total cost.

\subsection{Does a richer dispatch formulation change the coherence answer?}
\label{sec:discussion:richer}

Zhou et al.'s own separate-versus-joint ablation attributes up to 1.66
percentage points to joint or correlated forecasting in their two-stage,
reserve-based, distributionally robust dispatch, larger than the 0.64\%
ceiling we observe in our single-period newsvendor dispatch even under an
extreme stress test. Because their ablation is confounded with
marginal-accuracy changes (\Cref{sec:related-work}), we cannot conclude
from their result alone that a richer dispatch formulation genuinely
extracts more value from spatial correlation than ours does. We had
hypothesized that a dispatch with procured reserves and temporal coupling
might give the copula a wider channel through which to affect cost than
the narrow, curtailment-only channel we identify in
\Cref{sec:results:expb}.

We tested this hypothesis directly with a lightweight extension of the
same ablation design: a $T{=}4$-period rolling horizon, adding a generator
ramp limit and a minimum-upward-headroom reserve \emph{proxy} (not full
reserve procurement) to Experiment B's exact \textsc{Real}/\textsc{Indep}
comparison. Writing $\sched_0,\dots,\sched_{T-1}$ for the per-period
first-stage schedules, this extends \Cref{eq:two-stage} to
\begin{subequations}
\label{eq:multiperiod}
\begin{align}
\min_{\sched_0,\dots,\sched_{T-1}} & \ \sum_{k=0}^{T-1} \sum_{\zone}\!\left(a_\zone \sched_{k,\zone}^2 \!+\! b_\zone \sched_{k,\zone}\right) \nonumber\\
& +\, \sum_{k=0}^{T-1}\frac{1}{S}\sum_{s=1}^{S} \Big[\pshort\!\sum_{\zone}\! u_{k,\zone}^{(s)} \!+\! \vcurt\!\sum_{\zone}\!\mathrm{curt}_{k,\zone}^{(s)} \nonumber\\
&\hspace{5em} +\, \voll\!\sum_{\zone}\!\mathrm{shed}_{k,\zone}^{(s)}\Big] \label{eq:multiperiod:obj}\\
\text{s.t.}\ \ & \text{\Cref{eq:two-stage:bal,eq:two-stage:gmax,eq:two-stage:nonneg}} \nonumber\\
& \text{\Cref{eq:two-stage:flow} applied at each period } k, \nonumber\\
& \sched_{k,\zone} \le \Gmax{\zone}(1-\rho_{\text{res}}), \quad \forall k,\zone \label{eq:multiperiod:reserve}\\
& |\sched_{k,\zone} - \sched_{k-1,\zone}| \le \rho_{\text{ramp}} \Gmax{\zone}, \nonumber\\
&\hspace{5em} k=1,\dots,T{-}1,\ \forall \zone, \label{eq:multiperiod:ramp}
\end{align}
\end{subequations}
with ramp fraction $\rho_{\text{ramp}}=0.20$ and reserve-headroom fraction
$\rho_{\text{res}}=0.10$, chosen jointly, once, as a single here-and-now
decision against the whole horizon's ensemble (no intraday
re-optimization; each period draws its own $S$-scenario ensemble
independently, with no attempt to model temporal, as opposed to
cross-zone, dependence). Evaluated on the baseline and ceiling
configurations, pooled across 3 seeds ($n=600$ each), the result does not
support the wider-channel hypothesis: at baseline penetration the gain
moves from statistically indistinguishable from zero to a small but
detectable $0.03\%$ ($95\%$ CI $[0.02,0.05]$), while at the ceiling
configuration the gain \emph{falls} from $0.64\%$ to $0.41\%$ ($95\%$ CI
$[0.05,0.77]$), the opposite direction from what the hypothesis predicts.
A plausible reading is that the ramp limit, forcing the schedule to
change smoothly regardless of what the copula suggests, constrains
exactly the flexibility that would let correlation matter more; we did
not design this extension to isolate ramp from reserve, so we read this
as directional evidence, not a definitive answer: this modest form of
temporal coupling does not unlock a materially larger coherence value,
and both values remain far below the $\ge$1\% threshold.

These findings raise several open questions about the scope of our
central negative result, which we address directly in the following
limitations.

\section{Limitations}
\label{sec:limitations}

\textbf{The newsvendor dispatch ceiling.} The primary experiments use a
single-period newsvendor-style dispatch with no unit commitment, procured
reserves, ramp-rate limits, or temporal coupling across hours
(\Cref{sec:method:dispatch}); a full two-stage stochastic unit commitment
would delete the envelope-theorem gradient our decision-focused training
relies on and cost orders of magnitude more per solve. Every claim should
be read as characterizing this dispatch class, except
\Cref{sec:discussion:richer}'s four-period rolling-horizon extension,
which adds a ramp limit and a minimum-upward-headroom reserve
\emph{proxy} (not full reserve procurement) and does not train the
differentiable objective, so it is not constrained by the tradeoff above.

\textbf{Newsvendor cost parameters are mostly fixed.} $\vcurt=200$ and
$\voll=3000$ are fixed except for a sensitivity check on the coherence
ablation's ceiling configuration: sweeping $\voll$ from 1{,}000 to 9{,}000
changed the gain by less than $10^{-3}$ pp; sweeping $\vcurt$ from 100 to
600 moved the point estimate monotonically (0.72\% down to 0.34\%, versus
0.59\% at baseline 200; single-seed, $n{=}200$), staying well below the
$\ge$1\% threshold across this 6$\times$ range.

\textbf{A retracted mechanism.} We initially tried to demonstrate
coherence value via a differentiable loss committing a per-zone reserve
margin, $\text{committed}_z=\text{relu}(\mu_z-\kappa\sigma_z)$, using only
each zone's marginal mean/std; cross-zone correlation never entered this
expression and had an exactly-zero training gradient, so toggling the
generator between joint and independent modes measured a reserve-margin
artifact, not a coherence effect. We flag this as a pitfall for
differentiable coherence-modeling losses: verify the loss's gradient with
respect to dependence alone is nonzero, holding every marginal statistic
fixed. This is why we trust \Cref{sec:results:expb}'s ablation instead: it
asserts bit-identical marginals \emph{numerically} for every test
instance, and its nonzero effect in 7 of 10 configurations confirms it is
not similarly gradient-starved.

\textbf{Zhou et al.'s numbers are not independently re-verified.} The
comparison in \Cref{sec:discussion:zhou} relies on our own reading of
Zhou et al.~\cite{zhou2026decisionfocused}; re-verify against their
published artifacts before relying on it for a final venue decision.

\textbf{Three-seed pooling is modest.} Several Experiment B configurations
have confidence intervals that only narrowly exclude zero
(\Cref{fig:expb_forest}); additional seeds would tighten these.

\textbf{Bootstrap intervals treat test hours as exchangeable.} Our paired
bootstrap does not explicitly model autocorrelation between temporally
adjacent hours. A 24-hour moving-block bootstrap re-analysis of every
reported CI (no new experiments) found: for the decision-focused-training
claims (\Cref{sec:results:c1}), block-bootstrap intervals are
1.9-2.7$\times$ wider but every claim (baseline 3.31\%, 2.0$\times$
penetration 5.19\%, price-asymmetry 2.76\%, joint stress 4.55\%) still
excludes zero; for Experiment B (\Cref{sec:results:expb}), intervals are
nearly unchanged (0.86-1.05$\times$), with only 1 of 10 configurations
flipping zero-exclusion status, toward greater significance. Ignoring
temporal autocorrelation therefore does not appear to inflate any claim's
significance.

\textbf{No formal significance test for the skill-value gap.} The
$\lambda$-sweep (\Cref{sec:results:skillvalue}) is a single-seed,
descriptive curve without per-instance sidecar data for a paired CI; the
pattern is consistent with the better-powered decision-focused results
elsewhere but should not be read as carrying the same statistical weight.

\textbf{Decision-focused training costs substantially more compute.}
Wall-clock training time (not a controlled compute-profiling study) shows
decision-focused training taking roughly 5-40$\times$ longer than
accuracy-oriented training, widening on the larger CWE grid, consistent
with the per-solve overhead of differentiating through the dispatch
layer. Without a hardware-normalized comparison, this is a rough
order-of-magnitude caveat, not a precise cost-benefit figure.

\section{Conclusion}
\label{sec:conclusion}

We diagnosed what drives dispatch value in a single-period newsvendor-style
economic dispatch on real European grid data (a lightweight four-period,
ramp- and reserve-headroom-constrained extension, \Cref{sec:discussion:richer},
supports the same conclusion). The commonly assumed driver, spatial
coherence across renewable forecast sites, adds little value once cleanly
isolated from decision-focus and marginal-accuracy effects, is not
rescued by a finer scenario ensemble, and, once forecast errors are
stressed far beyond their historical scale, can be actively harmful when
approximated by the wrong parametric dependence model. Decision-focused
training, an established paradigm in this venue~\cite{zhang2026dflreview},
is significant and robust to adverse operating conditions in our
experiments.

These findings do not imply spatial dependence should be ignored. Rather,
a misspecified dependence model is indistinguishable from the
correctly-specified one at realistic forecast-error magnitudes but can
itself reduce operational value relative to no dependence once errors are
stressed well beyond those magnitudes (\Cref{sec:results:gausscop}),
arguing for evaluating dependence models by downstream decision
performance under realistic and stressed conditions alike, rather than
assuming greater statistical realism automatically improves dispatch.
Practitioners investing engineering effort in copula-based,
dependency-aware scenario generation should weigh that investment against
decision-focused training (noting its substantially higher training
compute cost, \Cref{sec:limitations}) and against decision-relevant
forecast improvements more broadly, not against marginal statistical
accuracy in isolation, since a 12\% energy-score improvement here changed
dispatch cost by less than 0.1\% (\Cref{sec:results:skillvalue}). We also provide a
strictly controlled, reusable ablation design for testing whether spatial
coherence matters in other dispatch formulations, since the confounded
separate-versus-joint comparisons common in prior work cannot answer this
question cleanly.

\textbf{Future work.} Concrete next steps, several of which were flagged
during this paper's own review process rather than deferred silently:
additional random seeds for Experiment B's borderline configurations; a
larger or synthetic network topology test (e.g.\ a DC-OPF or IEEE-118-bus
formulation with unit commitment, reserve procurement, and ramping,
beyond the lightweight four-period extension already reported in
\Cref{sec:discussion:richer}); a conditional (weather-regime- or
hour-of-day-stratified) dependence experiment using a modern
probabilistic forecaster rather than the persistence-based residual pool
used here; independent re-verification of Zhou et al.'s reported numbers
against their own artifacts; a per-instance cost-distribution analysis to
check whether gains concentrate in a tail of high-cost scenarios; and a
controlled, hardware-normalized computational-cost comparison (FLOPs and
solve time) to sharpen the rough estimate in \Cref{sec:limitations}.

\textbf{Data and code availability.} The full claim-to-file traceability
table appears in Appendix~A. All data are public (Open Power System
Data~\cite{wiese2018openpower}). Implementation and reproducibility
details are available from the authors upon reasonable request and
will be released in a public code repository upon publication.

\appendices
\section{Claim-to-File Mapping}
\label{app:claim-file-mapping}

Table~\ref{tab:claim-file-mapping} maps every numeric claim in this paper to
the exact raw data file(s) it was computed from. All files are
per-instance cost sidecars (\texttt{*\_percost.npy}) or run summaries
(\texttt{*.json}) produced directly by the experiment code, not
review-document summaries.

\begin{table}[t]
\centering
\tiny
\caption{Claim-to-file mapping. Paths are relative to the project's
\texttt{results/} directory unless otherwise prefixed (e.g.
\texttt{review-stage/}). Location abbreviations: R-coh = Results,
isolated-coherence subsection; R-skill = Results, skill-value
subsection; R-c1 = Results, decision-focused-value subsection;
R-scount = Results, scenario-count subsection; R-dep = Results,
dependence-model subsection; Meth = Method, data subsection; Disc =
Discussion, richer-formulations subsection; Lim = Limitations.}
\label{tab:claim-file-mapping}
\renewcommand{\arraystretch}{0.78}
\begin{tabular}{p{2.4cm}p{0.55cm}p{5.7cm}}
\toprule
\textbf{Claim} & \textbf{Loc.} & \textbf{Source file(s)} \\
\midrule
10-config coherence-gain forest plot, pooled + per-seed &
R-coh &
\seqsplit{expB\_corr\_only\_s\{0,1,2\}\_*.json} and matching \seqsplit{*\_percost.npy} \\
Curtailment identical to $<5\times10^{-4}$ pp across copula arms &
R-coh &
\seqsplit{expB\_corr\_only\_s\{0,1,2\}\_*.json} (\texttt{curt\_real\_pct}, \texttt{curt\_indep\_pct}; worst case \seqsplit{s1\_pen3.0\_rs2.0\_noclip.json}) \\
C2 skill-value gap: energy score 4.41$\to$3.87, cost 792.6-793.3 across $\lambda$ &
R-skill &
\seqsplit{cost\_s0\_lam\{0.0,0.05,0.1,0.3,1.0\}\_k1.0.json} \\
C1 penetration scaling: 3.31\% (1.0$\times$), 4.93\% (1.5$\times$), 5.19\% (2.0$\times$) &
R-c1 &
\seqsplit{\{crps,cost\}\_s\{0,1,2\}\_lam0.1\_k1.0\_cwe.json/.npy} (1.0$\times$);
\seqsplit{..\_cwe\_pen15\_uncap.json/.npy} (1.5$\times$);
\seqsplit{..\_cwe\_pen20\_uncap.json/.npy} (2.0$\times$) \\
Price asymmetry: 3.31\%$\to$2.76\% at $p_{\text{short}}$ 80$\to$160 &
R-c1 &
\seqsplit{\{crps,cost\}\_s\{0,1,2\}\_lam0.1\_k1.0\_cwe\_asym.json/.npy} \\
Joint S2$\times$S3 robustness: 4 corners, baseline/pen-only/price-only/JOINT &
R-c1 &
\seqsplit{\{crps,cost\}\_s\{0,1,2\}\_lam0.1\_k1.0\_\{cwe,cwe\_pen20\_uncap,cwe\_asym,cwe\_joint\}.json/.npy} \\
S1 interconnector congestion: 67-69\% of hours bind at baseline, 65-68\% at higher penetration &
Meth &
\seqsplit{crps\_s0\_lam0.1\_k1.0\_cwe.json} (\texttt{binding\_frac}, \texttt{binding\_per\_line}) \\
Scenario-count sensitivity: $S=20,50,100,200$, single-seed (4 configs); baseline and ceiling repeated 3-seed pooled &
R-scount &
\seqsplit{expB\_corr\_only\_s0\_*\_scount\{50,100,200\}.json}; \seqsplit{expB\_corr\_only\_s\{0,1,2\}\_pen1.0\_rs1.0\_scount\{20,50,100,200\}.json} (baseline); \seqsplit{expB\_corr\_only\_s\{0,1,2\}\_pen3.0\_rs8.0\_noclip\_scount\{20,50,100,200\}.json} (ceiling) \\
Dependence-model sensitivity: \textsc{Real}/\textsc{Gauss}/\textsc{Indep} at baseline, $2.0\times$, $3.0\times$, $8\times$-stress ceiling &
R-dep &
\seqsplit{expB\_gausscop\_s\{0,1,2\}\_pen1.0\_rs1.0\_baseline.json}; \seqsplit{..\_pen2.0\_rs1.0\_pen20.json}; \seqsplit{..\_pen3.0\_rs1.0\_pen30.json}; \seqsplit{..\_pen3.0\_rs8.0\_main.json}; matching \seqsplit{*\_percost.npy} \\
Multi-period (T=4) ramp/reserve-headroom extension: baseline and ceiling &
Disc &
\seqsplit{expB\_multiperiod\_s\{0,1,2\}\_pen1.0\_rs1.0\_T4\_main.json}; \seqsplit{..\_pen3.0\_rs8.0\_T4\_main.json} and matching \seqsplit{*\_percost.npy} \\
Moving-block bootstrap re-analysis: all 10 configs + 4 C1-family claims &
Lim &
\seqsplit{review-stage/block\_bootstrap\_reanalysis.json}, \seqsplit{review-stage/block\_bootstrap\_c1\_reanalysis.json} \\
Newsvendor cost-parameter sensitivity: $\voll\in[1000,9000]$, $\vcurt\in[100,600]$, ceiling config &
Lim &
\seqsplit{expB\_corr\_only\_s0\_pen1.0\_rs1.0\_voll\{1000,6000,9000\}.0\_vollsweep.json}; \seqsplit{..\_pen3.0\_rs8.0\_noclip\_voll\{1000,6000,9000\}.0\_vollsweep.json}; \seqsplit{..\_pen3.0\_rs8.0\_noclip\_vcurt\{100,400,600\}.0\_vcurtsweep.json} \\
Congestion (NTC) sensitivity: $\alpha_{\text{NTC}}\in\{0.05,0.15,0.20\}$ vs.\ $0.10$ baseline, near-zero + ceiling configs &
R-coh &
\seqsplit{expB\_corr\_only\_s\{0,1,2\}\_pen\{1.0,2.0,3.0\}\_rs\{1.0,8.0\}\_*\_ntc\{005,015,020\}.json} \\
\bottomrule
\end{tabular}
\end{table}

\footnotesize
\bibliographystyle{IEEEtran}
\bibliography{references}

\end{document}